\documentclass[preprint,aps,amssymb,amsmath,showpacs]{revtex4}
\usepackage{graphicx}
\usepackage{psfrag}
\usepackage{amssymb}
\newcommand{\be}{\begin{eqnarray}}
\newcommand{\ee}{\end{eqnarray}}

\begin{document}
%\title{Central potential problem with both Coulomb and harmonic oscillator potentials }
\title{ Series solution of a central potential problem with three-term recursion relation}
\author{ Jishnu Goswami, Chandan Mondal and Dipankar Chakrabarti}
%footnote{Email: dipankar@iitk.ac.in}}
\affiliation{Department of Physics, 
Indian Institute of Technology Kanpur,\\
Kanpur-208016,
India.}
\date{\today}

\begin{abstract}
The series solution of the radial part of  the Schr\"odinger equation for  simultaneous coulomb and harmonic potential  involves  three-term recursion relation and is thus difficult to solve for bound states. We have suggested a simple method  to solve for low lying states.  Finite polynomial solutions exist only if the coulomb and oscillator potentials are nontrivially related.

\end{abstract}
%\keywords{series solution, three-term recursion relation, truncation, coulomb potential, harmonic oscillator}
\maketitle

%% \linenumbers

%% main text%%%%%%%%%%%%%%%%%%%%%%%%%%%%%%%%%%%%%%%%%%%%%%%%%%%%%%%%%%%%%%

\section{Introduction}

In the quantum mechanics text books \cite{books},  the radial part of the Schr\"odinger equation for a central potential problem is solved by Frobenius' method where in general a  the two-term recursion relation relates one coefficient of the series to another one.  Normally one needs to truncate the series to a finite polynomial to have normalizable bound state wavefunctions. We also get the energy eigenvalues from the condition of  the series truncation.  But we find very rare comments about   the recursion relation  involving more than two coefficients as they are difficult to solve.  In the a popular  quantum mechanics book, in the context of hydrogen atom problem, the three-term recursion relation has been commented as ``enormously more difficult to work with" compared to a two-term recursion relation\cite{book2}. We can avoid the three-term recursion relation in hydrogen atom by choosing the functional form of the radial wavefunction from its behavior at small and large $r$.  Here we discuss    one such interesting example where we cannot avoid a three-term recursion relation.  Consider the Schr\"odinger equation in three dimensions when both coulomb and harmonic oscillator potentials are present together, i.e., the potential is given by
\be
V(r)=-\alpha\frac{1}{r}+\frac{1}{2} m\omega^2 r^2
\ee
(where $\alpha=e^2/(4\pi\epsilon_0)$ for hydrogen atom problem). 
The series solution of the radial part for this potential involves a three-term  recursion relation.
 For the bound state solution the series  needs to be truncated to a polynomial. The main difficulty of this problem is that 
the analytic solution for  any arbitrary energy level   in general form is very difficult to obtain.
In \cite{caho}, a similar type of problem was addressed with anisotropic harmonic oscillator with frequency ratio of the oscillators  in different directions chosen in  such a way that separation of variables in the parabolic co-ordinates works. If one takes isotropic oscillator, then it is not possible to make the separation of variables  for all three parabolic coordinates as done in \cite{caho}. The authors of that paper tried  a series solution which again resulted in the three-term recursion relation  which cannot be solved analytically and    they solved numerically in two dimensions for a special case when the coulomb term is zero.  Hall, Saad and Sen \cite{math} solved the isotropic case in d-dimensions  with more rigorous mathematical approach. They showed that to have polynomial solutions the parameters in the potential need to satisfy specific conditions depending on the order of the polynomial.

In this paper, we have suggested a very simplistic  method to obtain the low lying energy eigenvalues without much mathematical complications.  Converting the recursion relations involving three coefficients to relations involving only two, we can find out  the conditions for the series to be terminated to give normalizable bound state wavefunctions  and the energy eigenvalues.
% Our results are in exact agreement with the   solutions  for $d=3$ in \cite{math}.

\section{The radial equation}
We write the Schr\"dinger equation in spherical polar coordinates. After the separation of variables, the angular part of the Schr\"odinger equation can very easily be solved and can be found in any standard quantum mechanics book\cite{books}, the solution is given by the spherical harmonics $Y_{lm}(\theta,\phi)$. The radial equation is given by
\be
-\frac{\hbar^2}{2m}\frac{d^2u(r)}{dr^2}+\big[-\frac{\alpha}{r}+\frac{1}{2}m\omega^2 r^2
 +\frac{\hbar^2}{2m}\frac{l(l+1)}{r^2}\big]u(r)=Eu(r)\label{radial}
\ee
where $l$ is the azimuthal quantum number.
The effective potential 
\be
V_{eff}=-\frac{\alpha}{r}+\frac{1}{2}m\omega^2 r^2
 +\frac{\hbar^2}{2m}\frac{l(l+1)}{r^2}.
 \ee
%A schematic plot of the potential %for arbitrary $\alpha$ and $\omega$ 
%is shown in the Fig. \ref{fig1}. We have assumed that  strength of the coulomb potential is not too large to pull down the effective potential below zero, so that all the bound states will have positive energies.
% \begin{figure}[htp]
%  \begin{center}
 % \includegraphics[width=10cm]{V_eff.pdf}
%  \caption{\label{fig1} Effective potential (in arbitrary scale).}
%  \end{center}
%  \end{figure}  
Let us define 
\be 
k=\frac{\sqrt{2mE}}{\hbar},~~\rho=k r,~~\rho_0=\frac{2m\alpha}{\hbar^2 k},~~{\rm and}~ \rho_1=\frac{m\omega}{\hbar k^2}.\label{def}
\ee
Then the radial equation (Eq.\ref{radial}) can be written as
\be
\frac{d^2u}{d\rho^2}=\big[\rho_1^2\rho^2+\frac{l(l+1)}{\rho^2}-\frac{\rho_0}{\rho}-1\big]u\label{rad2}
\ee
The asymptotic behavior is   determined by
\be
\frac{d^2u}{d\rho^2}=\rho_1^2\rho^2 u
\ee
The general solution of this equation is given by the parabolic cylinder functions\cite{AS} $C_1 D_{-1/2}(\sqrt{2\rho_1}~\rho) + C_2 D_{-1/2}(i\sqrt{2\rho_1}~\rho)$  (where $C_1$ and $C_2$ are constants).
With the condition that the wavefunction vanishes at infinity,  the solution for $\rho\to\infty$  goes as 
\be
u\sim e^{-\rho_1\rho^2/2}
\ee
whereas the bevavior at small distances ($\rho\to 0$) is given by
%determined by
%\be
%\frac{d^2u}{d\rho^2}=\frac{l(l+1)}{\rho^2}u
%\ee
%and with the physical condition that the wavefunction is finite at the origin, the solution of the above equation is
\be
u\sim \rho^{l+1}
\ee
So, we assume that  the general form of radial wavefunction is given by
\be
u(\rho)=\rho^{l+1}e^{-\rho_1\rho^2/2} v(\rho)
\ee
With this substitution, Eq.(\ref{rad2}) becomes
\be
\rho\frac{d^2v}{d\rho^2}+2(l+1-\rho_1\rho^2)\frac{dv}{d\rho}+[\rho_0-\rho\{-1+(2l+3)\rho_1\}]v=0\label{radv}
\ee
%%%%%%%%%%%%%%%%%%
\subsection{series solution}
%%%%%%%%%%%%%%%%%
% $\rho=0$ is a regular singular point and it is possible to find at least one  series solution for the above eqnuation. 
%%%%%%%%%%%%%%%%%%%%%%%%%%
%Consider a second order differential equation of the form 
%\be 
%a x^2 y^{\prime\prime}+x A(x) y^\prime+B(x)y=0
%\ee
%with 
%\be
%A(x)=A_0+A_1 x+A_2 x^2=\cdots~~ {\rm and}~B(x)=B_0+B_1 x+B_2x^2+\cdots \nonumber
%\ee
%then the indicial equation is given by
%\be
%a k(k-1)+A_0 k+B_0=0
%\ee
%Multiplying the Eq.(\ref{radv}) with $\rho$ and comparing with the above  equations we get $a=1,~A_0=2(l+1),B_0=0$ and 
%%%%%%%%%%%%%%%%%%%%%%%%%%%%%%%
%The indicial equation  is given by
%\be k(k-1)+2(l+1)k=0 ~{\rm i.e.,}~k(k+2l+1)=0.
%\ee
% The roots of the indicial equations are $ k=0$,  $k=-(2l+1)$ and a series solution is possible for $k=0$. 
 %Since $l$ is an integer, the roots differ by an integer and it is possible to find a series solution for the highest root of the indicial eqn. Thus the series expansion of $v(\rho)$  for $k=0$ can be written as
Substituting 
\be 
v(\rho)= \sum_{i=0}^\infty c_i\rho^i
\ee
in  Eq.(\ref{radv}) we get
%\be
%\rho v^{\prime\prime}+2(l+1)v^{\prime}-2\rho_1\rho^2v^{\prime} +\rho_0 v-\rho[-1+(2l+3)\rho_1]v=0
%\ee
%\be
%\sum_{i=0}^\infty i(i+1)c_{i+1}\rho^i &+& 2(l+1)\sum_{i=0}^\infty (i+1)c_{i+1}\rho^i-2\rho_1\sum_{i=0}^\infty (i+1)c_{i+1}\rho^{i+2}\nonumber\\
%&+ &\rho_0 \sum_{i=0}^\infty c_i\rho^i-[(2l+3)\rho_1-1] \sum_{i=0}^\infty c_i\rho^{i+1}=0
%\ee
\be
\sum_{i=0}^\infty i(i+1)c_{i+1}\rho^i &+& 2(l+1)\sum_{i=0}^\infty (i+1)c_{i+1}\rho^i-2\rho_1\sum_{i=0}^\infty i c_{i}\rho^{i+1}\nonumber\\
&+ &\rho_0 \sum_{i=0}^\infty c_i\rho^i-[(2l+3)\rho_1-1] \sum_{i=0}^\infty c_i\rho^{i+1}=0
\ee
%%%%%%%%%%%%%%%%%%%
%The coefficient of $\rho^0$ gives the condition
%\be
%[2(l+1) c_1 +\rho_0 c_0] =0 ~~i.e., ~ c_1=-\frac{\rho_0}{2(l+1)}c_0.
%\ee
%%%%%%%%%%%%%%%%%%%
The coeffients $c_i$ satisifies the recursion relation 
\be
c_{i+1}=\frac{[a+2\rho_1(i-1)]c_{i-1}-\rho_0 c_i}{(i+1)(2 l+2+i)} \label{rec}
\ee
where $a=-1+(2l+3)\rho_1$ and
\be
 c_1=-\frac{\rho_0}{2(l+1)}c_0.
\ee
%Let us first look at the bevaviour of the solution for large $i$.  For very large values of $i$ we can approximate
%\be
%c_{i+1}= 2\rho_1\frac{ c_{i-1}}{i+1}
%=\left\{\begin{array}{ll}
%\frac{(2\rho_1)^{i/2}}{(i+1)!!}c_1 & {\rm if ~i~is~ even}\\
%\frac{(2\rho_1)^{(i+1)/2}}{(i+1)!!}c_0 & {\rm if ~i~is~ odd}
%\end{array}\right.
%\ee
Looking at the solution for large $i$, the approximate beviour of the series solution is given by
\be
 v(\rho)\sim c_o\sum_{i=1,3,5...}^\infty  \frac{\sqrt{2\rho_1}}{(i+1)!!} (\sqrt{2\rho_1}\rho)^i+c_1\sum_{i=2,4...}^\infty  \frac{1}{(i+1)!!} (\sqrt{2\rho_1}\rho)^i \ee
 Each of the terms in the above expression goes as $e^{\rho_1\rho^2}$ and thus spoils the asymptotic behavior of the radial wavefunction $u(r)$.  So,
 %Only if the coefficients $c_1$ and $c_0$ are such that the two terms in the above expression cancel each other, then we can expect to get a normalizable  wavefunction, otherwise 
   we need to truncate  the series. Note that here the recursion relation Eq.(\ref{rec}) involves three  coefficients $c_{i+1},~c_i, ~c_{i-1}$ and the truncation is very tricky.  If we want to truncate the series in the conventional way that is if we set the coefficient $c_{n+1}=0$ 
   %for some value of $i=n$,
    it does not guarantee the termination of the series.
   %  that the all terms for $i>n$ are zero and the series does not terminate. 
    On top of that, if we set 
 $c_{n+1}=0$ then it implies from Eq.(\ref{rec}) that 
 \be
 \frac{ c_n}{c_{n-1}}=\frac{a+2\rho_1(n-1)}{\rho_0}\label{div}
 \ee
 increases as $n$ increases i.e., the series does not converge which contradicts the original recursion relation! In its present form, we cannot set Eq.(\ref{rec}) to be zero.
% and  we need to stop the exponential behavior of the series for large $\rho$  in some other way.
 
 To truncate the series we need the recursion relation with only two coefficients. For this purpose we rewrite the first few coefficients in term of the lowest order coefficient $c_0$:
 \be
 c_1 &=& -\frac{\rho_0}{2(l+1)}c_0\nonumber\\
 c_2&=& \frac{a c_0-\rho_0 c_1}{2(2l+3)}=\frac{(2l+2) a+\rho_0^2}{2(2l+2)(2l+3)}c_0\nonumber\\
 c_3 &=& \frac{(a+2\rho_1) c_1-\rho_0 c_2}{3(2l+4)}=-\frac{\rho_0[2(2l+3)( a+2\rho_1) +(2(l+1)a+\rho_0^2)]}{6(2l+2)(2l+3)(2l+4)}c_0 \label{rec2}
 \ee
etc. Alternatively, one can also write down the coefficient $c_n$  in  terms of $c_{n-1}$ as 
\be
  c_1 &=& -\frac{\rho_0}{2(l+1)}c_0\nonumber\\
c_2&=& \frac{a c_0-\rho_0 c_1}{2(2l+3)}= -\frac{2(l+1)a+\rho_0^2}{2(2l+3)\rho_0}c_1\nonumber\\
 c_3 &=& \frac{(a+2\rho_1)c_1-\rho_0 c_2}{3(2l+4)}=-\frac{\rho_0[2(2l+3)( a+2\rho_1) +(2(l+1)a+\rho_0^2)]}{(2(l+1)a+\rho_0^2)3(2l+4)}c_2\label{rec2a}
 \ee
and so on. 
It is not possible to write down  the recursion relation for a general term $c_n$ in these forms,
% of Eq. (\ref{rec2}),
 but we can write down as many terms as we wish. So, in place of a general solution let us look at the low lying  solutions.
%Since $c_0\ne0$, it implies $c_1\ne 0$ also. 
The lowest possible term that can be set to zero is $c_2$  (this corresponds to $i=1$ in Eq.(\ref{rec})). If $c_2=0$ then $c_3=0$ only if the coefficient of $c_1$ in the recursion relation for $c_3$ i.e., $(a+2\rho_1)=0$  which puts an additional constraint on the energy eigenvalues.  
%Though the denominator in $c_3$ is  zero as $c_2=0$, there is no divergence. Once $c_2=c_3=0$  all other higher order coefficents $ c_4=c_5=\cdots=0$ and the series terminates.  When we terminate the series  at $i=n$ both  Eq.(\ref{rec2}) and Eq.({\ref{rec2a} )produce the same energy eigenvalues.
 The conditions that $c_2=0$ is given by 
\be
2(l+1)a+\rho_0^2=0.\label{E1}
\ee
Since $\rho_1=\frac{\hbar\omega}{2E}$ and $\rho_0^2=\frac{2m\alpha^2}{\hbar^2 E}$,  Eq.(\ref{E1}) gives the energy eigenvalues
\be
E_{1l}=(2l+3)\frac{1}{2}\hbar\omega+\frac{m\alpha^2}{(l+1)\hbar^2}.\label{e1l}
\ee
The condition that $c_3$ also be zero is given by 
\be
a+2\rho_1&=&0 \label{cond1}
\ee
which gives another expression for energy eigenvalue
\be
E_{1l}=\frac{1}{2}(2l+5)\hbar\omega.
\ee
Eq.(\ref{cond1})  together with Eq.(\ref{E1}) gives the condition  
\be
%(l+1)\hbar\omega=\frac{m\alpha^2}{\hbar^2} ~~i.e., ~~
\beta=(\frac{m\alpha^2}{\hbar^2})/(\hbar\omega)=l+1={\rm integer}
\ee
Thus, the infinite series can be terminated into a finite polynomial only if the parameters in the effective potential satisfy a nontrivial condition!  The above results   exactly agree with the results in \cite{math}.
%the extra condition relates the frequency of the harmonic oscillator with the strength of the coulomb potential in a nontrivial way.  
The minimum value of the ratio $\beta$ is one thus this method is not applicable  with  $\alpha=0$. 
With this condition the energy eigenvalue Eq.(\ref{e1l}) becomes
\be
 E_{1l}=(l+\frac{5}{2})\hbar\omega=\frac{3}{2}\hbar\omega+\frac{m\alpha^2}{\hbar^2}\label{gs}
 \ee
It is interesting to note that though we have never assumed any particular value for $l$,  the  final expression for the energy (Eq.(\ref{gs})) corresponds to $l=0$ in Eq.(\ref{e1l}).
Now,  let us set $c_3=0$ (i.e., $i=2$ in  Eq.(\ref{rec})) with the condition that $a+4\rho_1=0$ so that $c_4$ also becomes zero and the series terminates to a second order polynomial.  Then from Eqs.(\ref{rec2})   we get
\be
2(a+2\rho_1)(2l+3)+[2(l+1)a+\rho_0^2]=0 %(2l+4) a + 4\rho_1(2l+3)+\rho_0^2=0
\ee
which gives
\be
 E_{2l}=\frac{3(2l+3)(l+2)}{2(3l+4)}\hbar\omega+\frac{m\alpha^2}{(3l+4)\hbar^2}\label{e2}
 \ee
and the condition on the oscillator frequency and the strength of the coulomb potential  is given by the  dimensionless ratio 
\be
 \beta=4 l+5
%\frac{m\alpha^2}{\hbar^2}/(\hbar\omega)=4l+5
\ee
in agreement with \cite{math}. With this condition the energy eigenvalue reduces to
\be
 E_{2l}=(l+\frac{7}{2})\hbar\omega=\frac{9}{4}\hbar\omega+\frac{m\alpha^2}{4\hbar^2}
 \ee
 which again corresponds to $l=0$ in Eq.(\ref{e2}). 
  If we set the coefficient $c_n=0$ then  the series terminates if the pre-factor of $c_{n-1}$ in the recursion relation for $c_{n+1}$ is zero  i.e., 
  $ [a+2\rho_1(n-1)]=0$ which gives the energy eigenvaule 
  \be
  E_{nl}
  %&((n-1)+l+\frac{3}{2})\hbar\omega,~~n=2,3,4\cdots\nonumber\\
  =(n+l+\frac{3}{2})\hbar\omega, ~~n=1,2,3\cdots.
  \ee
%  which are exactly the eigenvalues of a three dimensional oscillator except the ground state.
  This condition ensures that the ratio $c_n/c_{n-1}$ in Eq.(\ref{div}) is zero and the divergence of the series does not arise.
  As the general condition for $c_n=0$ can not be written down, we cannot determine the condition amongst   $\omega$, $\alpha$ and $l$ for arbitrary $n$ .
  %%%%%%%%%%%%%%%
  \section{Conclusion}
  %%%%%%%%%%%%%%%
%  
 
 We have proposed a very simplistic method to extract the low lying eigenvalues by truncating a series involving three-term recursion relation. We have considered the quantum mechanical problem with  both harmonic oscillator and coulomb potentials. 
  The series solution of the radial equation results in a recursion relation involving three  coefficients and is difficult to  truncate the series into a polynomial.    We have shown  that the low lying eigenvalues and eigenfunctions can be  obtained by  a very simple method by writing the coefficient $c_n$ in terms of the lowest order coefficient $c_0$ or the preceding coefficient $c_{n-1}$.
  % If we write the recursion relations involving only $c_n$ and $c_{n-1}$ then also we get the same solutions.
  The series truncation needs an extra condition on the harmonic oscillator frequency and the coulomb strength. Depending on the order of the polynomial, the relation between $\omega$ and $\alpha$ comes out to be different, but  the dimensionless ratio $\beta$  always takes only integer values.  
   The energy eigenvalues can be written  as purely harmonic oscillator or hydrogenic energy levels  
   %completely in terms of $\hbar\omega$ or $(m\alpha^2/\hbar^2)$ 
   and the eigenvalues then depends on  angular momentum  $l$, on the other hand  the energies can also be written in combination  of both harmonic oscillator  and hydrogen atom energies by completely eliminating $l$ which is exactly same as the eigenvalue for $l=0$ without the additional constraint  between $\alpha$ and $\omega$.  Using our method we can obtain a general expression for the energy eigenvalue for terminating the series at any arbitrary  $n$-th order polynomial, but  we cannot evaluate  the relation between $\alpha$ and $\omega$ in the general form.  Thus this method provides a very simple way to extract the low lying eigenvalues when the recursion relation involves more than two coefficients.

  \end{document}